\documentstyle[12pt]{article}
\def\beq{\begin{equation}}
\def\brr{\begin{array}}
\def\err{\end{array}}
\def\eeq{\end{equation}}
\def\bea{\begin{eqnarray}}
\def\eea{\end{eqnarray}}

\def\tr{\mbox{Tr}\, }
\def\ni{\noindent}

\def\nn{\nonumber}
\def\ms{\medskip}

\begin{document}

\hfill HUPD-92-09

\hfill UB-ECM-PF 92/21

\hfill August 1992
\mbox{}

\vspace*{6mm}

\begin{center}

{\LARGE \bf
Laurent series representation for the open superstring free energy}

\vspace{8mm}

{\sc A.A. Bytsenko} \\ {\it Physico-Mechanical Faculty, State
Technical
University, St Petersburg 195251, Russia} \\
{\sc E. Elizalde} \\
{\it Department E.C.M., Faculty of Physics, University of
Barcelona,
Diagonal 647, 08028 Barcelona, Spain} \\
{\sc S.D. Odintsov}\footnote{On sabbatical leave from
Tomsk Pedagogical Institute, 634041 Tomsk, Russia.} \\
{\it Department of Physics, Faculty of Science, Hiroshima
University,
Higashi-Hiroshima 724, Japan} \\
 {\sc S. Zerbini} \\
{\it Department of Physics, University of Trento, 38050 Povo,
Italy}
\ms

\vspace{1cm}

{\bf Abstract}

\end{center}

Open superstrings at non-zero temperature are considered. A novel
representation for the free energy (Laurent series representation)
is
constructed. It is shown that the Hagedorn temperature arises in
this
formalism as the convergence condition (specifically, the radius of
convergence) of the Laurent series.

\vspace{1cm}

\noindent PACS: \begin{quote} 11.17 Theories of strings and other
extended objects, \ 03.70 Theory of quantized fields, \
04.50 Unified theories and other theories of gravitation.

 \end{quote}

\newpage

Superstring thermodynamics [1-4] has attracted considerable
attention recently (for a general review of strings at non-zero
temperature see [1]). One of the main motivations for this study is
connected with the fact that string thermodynamics can be relevant
for the description of the very early universe [5]. On the other
hand, the essential ingredient of string theory at non-zero
temperature is the famous Hagedorn temperature, above which the
free energy diverges.

According to the most popular viewpoint, the Hagedorn temperature
is the critical temperature of some phase transition to some new
phase (topological string theory? [6]), which describes string
theory above the Hagedorn temperature. Unfortunately, at present
there is very little to be said about this new phase of string
theory
at non-zero temperature. In  such a situation, it is reasonable to
make an additional effort in order to try to understand more deeply
string thermodynamics below the Hagedorn temperature, before
becoming
involved in the phase transition itself.

In particular, we still do not understand quite properly the
physical origin of the Hagedorn temperature or  string theory very
near the Hagedorn temperature [1-4]. Notice, however, the recent
very
interesting attempts [7,8] to use string theory at non-zero
temperature in the presence of world-sheet boundaries [8], for the
description of the high temperature limit of the confining phase of
large-n SU(n) Yang-Mills theory.

One can use different representations (in particular, connected
with different ensembles) for the calculation of the string free
energy. For example, the microcanonical ensemble representation for
the string free energy is more convenient than the canonical
ensemble representation to investigate string thermodynamics near
the Hagedorn temperature. However, all the well-known
representations for the free energy (see for example [1]) are
integral representations, in which the Hagedorn temperature appears
as the convergence condition in the ultraviolet limit. In order to
discuss the high-temperature or the low-temperature limits in such
representations one must expand the integral in the form of a
corresponding series. This is a difficult task.

In the present paper we construct a different (non-integral)
representation for the superstring free energy. This representation
is quite convenient both at high and at low  temperatures, because
it is given in terms of a Laurent series. What is more interesting,
the Hagedorn temperature appears in this formalism as the radius of
convergence of the Laurent series.

Let us start with the discussion of the free energy in field theory
at non-zero temperature. It is quite well-known that the one-loop
free energy for the bosonic (b) or fermionic (f) degree of freedom
in d-dimensional space is given by
\beq
F_{b,f} = \pm \frac{1}{\beta} \int \frac{d^{d-1} k}{(2\pi )^{d-1}}
\log \left(1\mp  e^{-\beta u_k} \right)
\eeq
where $\beta$ is the inverse temperature, $u_k=\sqrt{k^2+m^2}$, and
$m$ is the mass for the corresponding degree of freedom.
Expanding the logarithm and performing the (elementary) integration
one easily gets (see for example [9])
\bea
F_b &=& - \sum_{n=1}^{\infty} (\beta n)^{-d/2} \pi^{-d/2} 2^{1-d/2}
m^{d/2} K_{d/2} (\beta n m), \nn \\
F_f &=& - \sum_{n=1}^{\infty} (-1)^n (\beta n)^{-d/2} \pi^{-d/2}
2^{1-d/2} m^{d/2} K_{d/2} (\beta n m),
\eea
where $ K_{d/2} (z)$ are the modified Bessel functions. Using the
integral representation for the Bessel function
\beq
 K_{d/2} (z) = \frac{1}{2} \left( \frac{z}{2} \right)^{d/2}
\int_0^{\infty} ds \, s^{-1-d/2} e^{-s-z^2/(4s)},
\eeq
one can obtain the well-known proper time representation for the
one-loop free energy:
\beq
F_{b,f} = - \int_0^{\infty} ds \, \pi^{-d/2} 2^{-1-d/2}  s^{-1-d/2}
e^{-m^2s/2} \times \left\{ \brr{l} \left[ \theta_3  \left( 0 |
\frac{i\beta^2}{2\pi s} \right) -1 \right], \\
\left] 1-\theta_4  \left( 0 | \frac{i\beta^2}{2\pi s} \right)
\right].  \err \right.
\eeq
Expression (4) is usually the starting point  for the calculation
of the string free energy in the canonical ensemble (then $m^2$ is
the mass operator and for closed strings the corresponding
constraint  should be taken into account).

Let us now consider a different representation (in terms of a
Laurent series) for the bosonic or fermionic field free energy in
a
$d$-dimensional space-time. In field theory such a representation
has been introduced in refs. [10,11]. We start from the obvious
identity
\beq
\int d^d k \, f(k) = \frac{2\pi^{d/2}}{\Gamma (d/2)} \int dk \,
k^{d-1} f(k).
\eeq
Using it and upon integration of (1) by parts, we obtain
\beq
F_{b,f} =- \frac{(4\pi)^{(1-d)/2}}{(d-1)\Gamma ((d-1)/2)} \int dk^2
\, k^{d-1} \frac{1}{u_k \left( e^{\beta u_k} \mp 1 \right)}.
\eeq
For the factor in the integrand, $\left( e^{\beta u_k} \mp 1
\right)^{-1}$, we shall use the Mellin transform in the following
form [12]
\beq
 \frac{1}{ e^{ax} \mp 1 } = \frac{1}{2\pi i} \int_{c-
i\infty}^{c+i\infty} ds \,  \zeta^{(\mp)} (s) \Gamma (s)
(ax)^{-s},
\eeq
where Re $s=c, \ c>1$, for bosons and $c>0$ for fermions, $
\zeta^{(-)} (s)=\zeta (s)$ is the Riemann-Hurwitz zeta function and
$ \zeta^{(+)} (s)=\left( 1-2^{1-s} \right) \zeta (s)$ $=
\sum_{n=1}^{\infty} (-1)^{n-1} n^{-s}$, Re $s>0$.

Substituting (7) into (6), we get
\beq
F_{b,f} =- \frac{(4\pi)^{(1-d)/2}}{(d-1)\Gamma ((d-1)/2)} \int dk^2
\, k^{d-1} \frac{1}{2\pi i} \int_{c-i\infty}^{c+i\infty} ds \,
\zeta^{(\mp)} (s) \Gamma (s) \beta^{-s} u_k^{-1-s}.
\eeq
Integrating over $k$ with the help of the Euler beta function
$B(x,y)=\Gamma (x)\Gamma (y)/\Gamma (x+y)$ (notice that, owing to
absolute convergence, the order of integration over $k$ and $s$ can
be interchanged), we obtain
\beq
 \int dk^2 \, k^{d-1}  u_k^{-1-s} = (m^2)^{(d-s)/2} B \left(
\frac{d+1}{2}, \frac{s-d}{2} \right) = (m^2)^{(d-s)/2} \frac{
\Gamma \left( \frac{d+1}{2} \right) \Gamma \left( \frac{s-d}{2}
\right)}{ \Gamma \left( \frac{s+1}{2} \right)}, \ \ \mbox{Re} \, s
>d.
\eeq
Finally, one gets [10]
\beq
F_{b,f} =- 2^{-d} \pi^{(1-d)/2} \frac{1}{2\pi i} \int_{c-
i\infty}^{c+i\infty} ds \,  \Gamma (s) \frac{ \Gamma \left(
\frac{s-d}{2} \right)}{ \Gamma \left( \frac{s+1}{2} \right)}
\zeta^{(\mp)} (s)  \beta^{-s}  (m^2)^{(d-s)/2},  \ \ \mbox{Re} \,
s >d.
\eeq
This is the main expression which will be used for the calculation
of the one-loop string free energy. Notice that the new formula
(10)
is quite different from the usual one (4).

For string theory one can write the representation (10) in the form
\bea
F_{\mbox{bosonic string}} &=&- 2^{-d} \pi^{(1-d)/2} \frac{1}{2\pi
i} \int_{c-i\infty}^{c+i\infty} ds \,  \Gamma (s) \frac{
\Gamma \left( \frac{s-d}{2} \right)}{ \Gamma \left( \frac{s+1}{2}
\right)} \zeta (s)  \beta^{-s} \tr (M^2)^{(d-s)/2}, \nn \\  & &   d=26,
 \\
F_{\mbox{superstring}} &=&- 2^{-d+1} \pi^{(1-d)/2} \frac{1}{2\pi i}
\int_{c-i\infty}^{c+i\infty} ds \,  \Gamma (s) \frac{ \Gamma
\left( \frac{s-d}{2} \right)}{ \Gamma \left( \frac{s+1}{2} \right)}
\zeta (s)  (1-2^{-s}) \beta^{-s} \mbox{Str} \, (M^2)^{(d-s)/2}, \nn \\
& &
d=10. \nn  \eea
Here $M^2$ is the mass operator and for closed strings the
constraint should be introduced via the usual identity [1]. It is
interesting to observe that the fermionic contribution to the free
energy can be obtained from the bosonic one with the help of (10).
The closed result is [13]
\beq
F_f (\beta ) = - \left( 2F_b (2\beta ) -F_b (\beta ) \right).
\eeq

For bosonic strings the mass operator contains both infrared (due
to the presence of the tachyon in the spectrum) and ultraviolet
divergences, while for superstrings it contains only ultraviolet
divergences. Hence, the consideration of superstrings will be much
simpler from a technical point of view.

In what follows we shall consider open superstrings only, leaving
the (more involved) discussion of closed superstrings and bosonic
strings to another work. For the open superstring (without gauge
group) the spectrum is given by (see for example [14])
\beq
M^2=2\sum_{i=1}^{d-2}\sum_{n=1}^{\infty}  n \left( N_{ni}^b +
N_{ni}^f \right), \ \ \ \ d=10.
\eeq
In order to calculate Str $(M^2)^{(d-s)/2}$ it is convenient to use
the heat-kernel representation
\beq
\mbox{Str} \, (M^2)^{(d-s)/2}= \frac{1}{\Gamma \left( \frac{s-d}{2}
\right)} \int_0^{\infty} dt \, t^{ \frac{s-d}{2} -1} \, \mbox{Str}
\,
e^{-tM^2}.
\eeq
For $d=10$, it is known that
\beq
 \mbox{Str} \, e^{-tM^2} =\prod_{n=1}^{\infty} \left( \frac{1-e^{-
2tn}}{1+e^{-2tn}} \right)^{-8} = \left[ \theta_4 \left( 0 |
e^{-2t} \right)\right]^{-8}.
\eeq
It follows that
\beq
\left. \mbox{Str} \, (M^2)^{ \frac{d-s}{2}}\right|_{d=10}  =
\frac{2^{5-s/2}}{\Gamma \left( \frac{s}{2}-5 \right)}
\int_0^{\infty} dt \, t^{ \frac{s}{2} -6} \left[ \theta_4 \left( 0
| e^{-t} \right)\right]^{-8}.
\eeq
According to (10), Re $s>10$. This is why, for $s\leq 10$,
expression (16) contains the ultraviolet divergences ($t\rightarrow
0$).

Let us now analytically continue the integral (16) to the complex
$s$-plane. Recall that
\bea
\theta_4 \left( 0 | e^{-t} \right) &=&
\sqrt{\frac{\pi}{t}} \,
\theta_2 \left( 0 | e^{-\pi^2/t} \right) =
\sqrt{\frac{\pi}{t}}
\sum_{n=-\infty}^{+\infty} \exp \left[ -\frac{\pi^2}{t} \left( n-
\frac{1}{2} \right)^2 \right] \nn \\
&=& 2
\sqrt{\frac{\pi}{t}}
 \, e^{-\pi^2/(4t)} \left( 1+ e^{-9\pi^2/t}+ e^{-25\pi^2/(4t)} +
\cdots \right), \ \ \ \ t\longrightarrow 0,
\eea
and therefore
\beq
\left. \left[ \theta_4 \left( 0 | e^{-t} \right)\right]^{-
8} \right|_{t\rightarrow 0} = \frac{t^4}{2^8\pi^4} \, e^{2\pi^2/t}
-\frac{t^4}{2^5\pi^4} + {\cal O} \left( e^{-2\pi^2/t} \right).
\eeq
Hence, one can identically regularize the integral (16) in the
following way
\bea
 \mbox{Str} \, (M^2)^{5- \frac{s}{2}}& =&  \frac{2^{5-s/2}}{\Gamma
\left( \frac{s}{2}-5 \right)} \left\{ \int_0^{\infty} dt \, t^{
\frac{s}{2} -6} \left[ \left[ \theta_4 \left( 0 | e^{-t}
\right)\right]^{-8} - \frac{t^4}{2^8\pi^4} \, \left( e^{2\pi^2/t} -
8 \right) \right] \right. \nn \\
& +&\left.   \frac{1}{2^8\pi^4} \, \int_0^{\infty} dt \, t^{
\frac{s}{2} -2} \left( e^{2\pi^2/t} -8 \right)
\right\}.
\eea
Since the regularization of the integral $ \int_0^{\infty} dx \,
x^{\lambda}$ as an analytical function of $\lambda$ gives $
\int_0^{\infty} dx \, x^{\lambda} =0$, it turns out that the last
integral in (18) is equal to zero. Moreover, we have
\beq
 \int_0^{\infty} dt \, t^{ \frac{s}{2} -6} t^4 e^{2\pi^2/t}
=   \int_0^{\infty} dt \, t^{\left(- \frac{s}{2}
+1\right)-1}  e^{2\pi^2t} =(- 2\pi^2)^{ \frac{s}{2} -1} \Gamma
\left(1-\frac{s}{2} \right),
\eeq
and therefore
\beq
 \mbox{Str} \, (M^2)^{5- \frac{s}{2}} = \frac{2^{-4}\pi^{-
6}}{\Gamma \left(
\frac{s}{2}-5 \right)} \left[ \pi^s \Gamma \left(1-\frac{s}{2}
\right) \mbox{Re} \,  (-1)^{\frac{s}{2} -1}  +2 G(s,\mu)
\right],
\eeq
where
\beq
G(s, \mu)= \pi \left( \frac{\pi}{2} \right)^{s/2}  \int_0^{\mu} dt
\, t^{ \frac{s}{2} -6}
\left\{ \left[ \frac{1}{2} \theta_4 \left( 0 | e^{-\pi t}
\right)\right]^{- 8} -  t^4\left( e^{2\pi/t}-8 \right) \right\}.
\eeq
In (22) the infrared cutoff parameter $\mu$  has been introduced.
Such a regularization is necessary  for $s\geq 4$. On the next
stage of our calculations this regularization will be removed ($\mu
\rightarrow \infty$).

For the one-loop free energy, we obtain
\bea
F_{\mbox{superstring}} &=& - 2^{-13} \pi^{-21/2} \frac{1}{2\pi i}
\int_{c-i\infty}^{c+i\infty} ds \, \frac{ \Gamma
( s)}{ \Gamma \left( \frac{s+1}{2} \right)} \zeta (s) (1-2^{-s})
\beta^{-s}\nn \\
&\times& \left[ \pi^s \Gamma \left(1-\frac{s}{2} \right) \mbox{Re}
\,  (-1)^{\frac{s}{2} -1}  +2 G(s,\mu)
\right] \\
& \equiv &  - 2^{-13} \pi^{-21/2} \frac{1}{2\pi i}
\int_{c-i\infty}^{c+i\infty} ds \, \left[ \varphi (s) + \psi (s)
\right]. \nn
\eea
Here
\bea
\varphi (s) &=& (1-2^{-s} )  \mbox{Re} \,  (-1)^{\frac{s}{2} -1}
\frac{ \Gamma ( s) \Gamma \left(1- \frac{s}{2} \right)}{ \Gamma
\left( \frac{s+1}{2} \right)} \zeta (s) \left( \frac{\beta}{\pi}
\right)^{-s}, \\
\psi (s) &=& 2 (1-2^{-s} )  G(s,\mu) \,  \frac{ \Gamma ( s)}{
\Gamma \left( \frac{s+1}{2} \right)} \zeta (s)  \beta^{-s}.
\eea

The meromorphic function $\varphi (s)$ has first order poles at
$s=1$ and $s=2k$, $k=1,2, \ldots$. The meromorphic function $\psi
(s)$ has first order poles at $s=1$. The corresponding residues are
\beq
\mbox{Res} \, (\varphi (s),s=2k)=  (1-2^{-2k}) \frac{ \Gamma
(2k)\zeta (2k)}{ \Gamma \left(k+ \frac{1}{2} \right) \Gamma (k)}
\left( \frac{\beta}{\pi} \right)^{-2k}
\eeq
and
\beq
\mbox{Res} \, (\psi (s),s=1)=   G(1, \infty) \beta^{-1}.
\eeq
The pole of $\varphi$ for $s=1$  is also of first order but its
residue
is imaginary. There
are no
further poles contributing to the integral. One can see that the
regularization cutoff in the infrared domain is removed
automatically.

Finally, we obtain
\bea
F_{\mbox{superstring}} & =& -2^{-13} \pi^{-21/2} \left[ \sum_k
\mbox{Res} \, (\varphi (s),s=2k) +\mbox{Res} \, (\psi (s),s=1)
\right] \nn \\
&=& -2^{-13} \pi^{-21/2}  \left[
\sum_{k=1}^{\infty} A(k) \beta^{-2k} +G(1, \infty) \beta^{-1}
\right],
\eea
with
\beq
A(k)= (-1)^{k+1} (2^{2k}-1)\pi^{4k} \frac{ B_{2k}}{4 \Gamma
\left(k+\frac{1}{2} \right) \Gamma (k+1)},
\eeq
where $B_{2k}$ are the Bernoulli numbers.

Eq. (28) is the main result of this paper. It gives the Laurent
series representation of the open superstring free energy. This
representation is very convenient, both for low temperature and for
high-temperature calculations. In fact, the only thing to do in any
case is to take the neccessary amount of relevant terms of the
series (28). Notice, once more, that this series  is not just an
expansion: it actually provides the {\it exact} result for any
value of
$\beta$ for which the series exists.

Let us discuss this point in more detail, namely the convergence
condition
for the Laurent series (28). The usual convergence criterion (of
quocients) reads
 \beq
\lim_{k\rightarrow \infty} \frac{A(k+1)}{A(k)} = 4\pi^2 \beta^{-2}
\equiv q.
\eeq
For $q<1$ the series is convergent. Hence, expression (28) is
convergent
when $\beta > \beta_c =2\pi$, the Hagedorn temperature [1-4]. That
is, the Hagedorn temperature provides the convergence radius of the
above series, and vice-versa. It is interesting to notice that the
coefficients of the main part of the Laurent series (28) are
defined in the following way
\beq
A(k)= \frac{1 }{2\pi i} \oint_{\cal C} d\beta \, \beta^{k-1}
F(\beta),
\eeq
where ${\cal C}$ is an arbitrary, closed integration path interior
to the convergence circle of the series. The radius of this circle
of convergence corresponds to the critical temperature.

To summarize, in the present paper we have developed a formalism in
which the string free energy is expressed as a Laurent series. The
advantages of this formalism are appealing. To begin with, our
representation
is exact. Furthermore, it is very convenient both at low and at
high
(near
Hagedorn) temperatures, because (as is clear) one can easily
estimate
$F(\beta)$ from (28) to any required accuracy. Due to technical
reasons, here we have limited ourselves to the case of open
superstrings. The Laurent series representation for bosonic strings
and for closed superstrings will be discussed in another work.
\vspace{5mm}

\ni{\large \bf Acknowledgments}

S.D.O. acknowledges a grant from the Japan Society for the
Promotion of Science, which made his participation in this work
possible. S.D.O. wishes to thank Prof. T. Muta and the members of
the Particle Group at Hiroshima University for kind hospitality.
E.E. has been supported by DGICYT (Spain), research project
PB90-0022.

\newpage
%\renewcommand
%\baselinestretch{1.05}

%{\small

%}


\begin{thebibliography}{99}

\bibitem{} S.D. Odintsov, {\sl Rivista Nuovo Cim.} {\bf 15} (1992)
1.

\bibitem{} R. Rohm, {\sl Nucl. Phys.} {\bf B237} (1984) 553; K.
Kikkawa and M. Yamasaki, {\sl Phys. Lett.} {\bf B149} (1984) 357;
M. Bowick and L.C.R. Wijewardhana,  {\sl Phys. Rev. Lett.} {\bf 54}
(1985) 2485; M. Gleiser and J.G. Taylor,  {\sl Phys. Lett.} {\bf
B164}
(1985) 36; J.J. Atick and E. Witten, {\sl Nucl. Phys.} {\bf B310}
(1988) 291; S.D. Odintsov, {\sl Europhys. Lett.} {\bf 8} (1989)
207.

\bibitem{} E. Alvarez and M. Osorio, {\sl Phys. Rev.} {\bf D36}
(1987) 1175; M. McGuigan, {\sl Phys. Rev.} {\bf D38} (1988) 552.

\bibitem{} I. Antoniadis, J. Ellis and D.V. Nanopoulos, {\sl Phys.
Lett.} {\bf B199} (1987) 402; F. Englert and J. Orloff, {\sl Nucl.
Phys.} {\bf B334} (1990) 472; I.M. Lichtzier and S.D. Odintsov,
{\sl Yad. Fiz. (Sov. J. Nucl. Phys.)}
{\bf 51} (1990) 1473;ibid., {\sl Izw. V.U. Zov. (Sov. Phys.
Journal)}
{\bf 12} (1991) 3; I. Antoniadis and C. Kounnas,  {\sl Phys. Lett.}
{\bf B261} (1991) 369.

\bibitem{} A.A. Tseytlin and C. Vafa, {\sl Nucl. Phys.} {\bf
B372} (1992) 443; H. Cateau and K. Sumiyashi, Tokyo preprint, HEPTH
9205096, 1992; R. Brandenberger and C. Vafa, {\sl Nucl. Phys.} {\bf
B316} (1989) 391; A.A. Tseytlin, {\sl Class. Quant. Grav.} {\bf
9} (1992) 979.

\bibitem{} E. Witten, {\sl Commun. Math. Phys.} {\bf 117} (1988)
353; ibid. preprint IASSNS-HEP-89/66, 1989.

\bibitem{} J. Polchinski, {\sl Phys. Rev. Lett.} {\bf 68} (1992)
1267.

\bibitem{} M.B. Green, {\sl Phys. Lett.} {\bf B282} (1992) 380.

\bibitem{} E. Elizalde and A. Romeo, {\sl Phys. Rev.} {\bf D40}
(1989) 436; ibid. {\sl Int. J. Mod. Phys. A}, to appear.

\bibitem{} A.A. Bytsenko, L. Vanzo and S. Zerbini, preprint Trento
Univ. 259, 1992, to appear in {\sl Int. J. Mod. Phys. A}.

\bibitem{} A.A. Bytsenko, L. Vanzo and S. Zerbini, preprint Trento
Univ. 260, 1992, to appear in {\sl Phys. Lett. B}.

\bibitem{} A. Erdlyi, ed., {\sl Higher Transcendental Functions},
vols. 1 and 2, McGraw-Hill, New York, 1953.

\bibitem{} J.S. Dowker and J.P. Schofield,  {\sl Nucl. Phys.} {\bf
B327} (1989) 267.

\bibitem{} M.B. Green, J.H. Schwarz and E. Witten, {\sl Superstring
Theory}, Cambridge Univ. Press, Cambridge, 1987.
\end{thebibliography}
\end{document}